\documentclass[twocolumn,noshowpacs,nofootinbib,10pt]{revtex4}

\usepackage{float,xcolor,upgreek,yfonts,tikz}
\usepackage{color} 
\usepackage{graphicx}
\usepackage{graphicx,float}\usepackage[all]{xy}
\usepackage{amsmath,upgreek,accents}
\definecolor{applegreen}{rgb}{0.55, 0.71, 0.0}

\definecolor{amber}{rgb}{1.0, 0.55, 0.1}

\usepackage{amssymb}

\usepackage{alphabeta}
\usepackage{textgreek}
\newcommand{\beq}{\begin{eqnarray}}

\newcommand{\eeq}{\end{eqnarray}}
\newcommand{\be}{\begin{equation}}
\newcommand{\ee}{\end{equation}}

\newcommand{\bea}{\begin{eqnarray}}
\newcommand{\eea}{\end{eqnarray}}

\newcommand{\ba}{\begin{eqnarray}}
\newcommand{\ea}{\end{eqnarray}}
\newcommand{\clt}{\textcolor{black}}

\usepackage[T3,T1]{fontenc}

\usepackage[colorlinks,hyperindex,unicode]{hyperref}
\definecolor{green1}{RGB}{0,128,0} 
\hypersetup{hidelinks,backref=true,pagebackref=true,hyperindex=true,colorlinks=true,breaklinks=true,urlcolor= blue}
\hypersetup{%
  colorlinks = true,
  linkcolor  = blue,
  citecolor = cyan,
}
\usepackage{bookmark,textgreek}
\usepackage{hyperref,color,xcolor}

\newcommand\orcidroldao{{\href{https://orcid.org/0000-0003-3978-532X}{\orcidicon}}}
\newcommand\orcidwillians{{\href{https://orcid.org/0000-0001-9750-2637}{\orcidicon}}}
\newcommand{\orcidicon}{%
	\begin{tikzpicture}
	\draw[lime, fill=lime] (0,0)
		circle [radius=0.16]
		node[white] {{\fontfamily{qag}\selectfont \tiny ID}};
	\draw[white, fill=white] (-0.0625,0.095)
		circle [radius=0.007];
	\end{tikzpicture}	\hspace{-2mm}
}

\begin{document}
\title{Gravitational collapse in AdS: instabilities, 
turbulence, and information}

\author{W. Barreto\orcidwillians\!\!}
\affiliation{Federal University of ABC, Center of Natural Sciences, Santo Andr\'e, 09580-210, Brazil}\email{willians.barreto@ufabc.edu.br}
\affiliation{Centro de F\'{i}sica Fundamental, Universidad de Los Andes, M\'{e}rida 5101, Venezuela}
\author{R. da Rocha\orcidroldao\!\!}
\affiliation{Federal University of ABC, Center of Mathematics,  Santo Andr\'e, 09210-580, Brazil.}
\email{roldao.rocha@ufabc.edu.br}
\begin{abstract}The gravitational collapse in asymptotically AdS spacetimes is studied, using the differential configurational entropy (DCE). The DCE is shown to comply with the instability of the Einstein--Klein--Gordon system, leading to different stages in the route to turbulence and the final black
hole formation, including near the critical behavior. 
\end{abstract}
\maketitle
\section{Introduction}

Shannon introduced information theory for memoryless channels and sources in arbitrary communication systems. Information sources are Markov stochastic processes that encompass conditional entropy. Information entropy shares ergodicity and   Shannon used the ergodic theorem to characterize the optimal performance attainable when communicating information sources across constrained random media. The ergodic theorem was
introduced by defining a mathematical measure of the entropy or information in random mechanisms, also posing its asymptotic behavior \cite{Shannon:1948zz}.  Configurational entropy (CE)  designates a theory of data compression and source coding, encompassing transmission and storing information with optimal channel coding.
The main gist of the CE consists of the quantification of the amount of information in probability distributions describing physical systems. The CE has been comprehensively employed in machine learning and measures how much surprise there is in an event. Rare events thereby correspond to low probabilities and are more surprising, having more information than common events \cite{gs12a}. 
Therefore realizing that an unlikely event came about is more informative than acquiring the knowledge that a likely event has occurred.
Rare events are more uncertain or more surprising,  requesting more information to portray them than ordinary common events.
The CE has been thrivingly applied to investigate a large spectrum of relevant theories and physical models, proposing new features and corroborating to 
well-established physical phenomena \cite{gs12b}.

The CE relies on the same fundaments of Gibbs--Boltzmann statistical mechanics entropy, reporting the degree of disorder in discrete physical systems. Going to the continuum mechanical limit,  the differential configurational entropy (DCE) is the quantity that best quantifies the entropy of information \cite{Gleiser:2018kbq,Gleiser:2018jpd,Bernardini:2016hvx}. 
In any physical system, critical points of the DCE indicate which microstates have maximal stability of configuration, corresponding to the prevalent state that the system occupies. Indeed it regards a probability
distribution permitting the highest remaining uncertainty, corresponding to 
the maximum entropy principle. The DCE has been applied to the solution of a variety of relevant physical systems. 
Gauge/gravity dualities, black holes, and phase transitions were successfully investigated with the tools of the DCE. 
The first outstanding result regulated by the DCE in this context was the identification of the Hawking--Page transition as a critical instability measured by the DCE that underlies AdS (anti-de Sitter) black holes, splitting with pure radiation a state of thermal equilibrium in AdS \cite{Braga:2016wzx}, which is broken at a critical temperature also determined by the DCE \cite{Lee:2021rag}. Ref. \cite{Braga:2020opg} showed that the DCE is a logarithm function of the temperature of the plasma phase in the dual QCD, establishing a new link between the DCE and thermodynamics of AdS black holes.
Ref. \cite{Braga:2017fsb} used the DCE to 
study quantitative aspects of charmonia and bottomonia in AdS/QCD. Quarkonia dissociation processes in the quark-gluon plasma are correlated to a latent instability driven by the DCE \cite{Braga:2020hhs}.
Ref. \cite{Braga:2020myi} demonstrated that the DCE regulates the stability of heavy vector mesonic states against dissociation, in a thermal medium. 
Additionally, new aspects of AdS black holes have been paved, including the way information is stored in AdS black holes, in the context of the Hawking--Page phase transition and the DCE  \cite{Braga:2019jqg}. Second-order phase transitions regulated by the DCE and thermodynamical features of AdS black holes were established in Ref.  \cite{Lee:2017ero}. 
A variety of stellar distributions were explored in the DCE setup, % \cite{Fernandes-Silva:2019fez}, 
pointing toward the Chandrasekhar instability related to global minima of the DCE \cite{Gleiser:2013mga,Gleiser:2015rwa}.
%, including stellar solutions of the low-energy limit in string theory% \cite{dr21}. 
Ref. \cite{Bernardini:2019stn} evinced the DCE controlling particle and nuclear interactions, also in cosmological scales.
The mass spectra of hadronic states in several meson families were derived using DCE-type Regge trajectories in the soft- and hard-wall AdS/QCD models \cite{Bernardini:2018uuy,Ferreira:2019inu,daRocha:2021ntm,daRocha:2021imz}, 
%, also including families of baryons at finite temperature \cite{Ferreira:2020iry} 
as well as the prediction of the next generation of hadronic molecules, the hadrocharmonium, and the hybrid heavy-quark exotica in QCD \cite{Karapetyan:2021ufz}.
The DCE critical points complied with several other mesonic properties in the context of the holographic light-front QCD in Refs. \cite{Karapetyan:2018yhm,Karapetyan:2018oye,Karapetyan:2021vyh}, and played a relevant role in deriving the Higgs boson mass and its decays \cite{Alves:2020cmr}. 
Also, the DCE was employed to analyze topological field theories \cite{Bazeia:2018uyg,Bazeia:2021stz}.

\clt{The DCE was originally presented as a quantity that sets side by side  solutions of soliton-type to any approximation that describes it, removing degeneracies of ans\"atze with the same energy, when one regards models with a 
scalar field.  Ref. \cite{gs12a} showed that the higher the energy of a trial scalar field estimating exact solutions, the higher the DCE is, which underlies the system. The DCE, therefore, determines the best fit for solutions involving scalar fields.  Ref. \cite{Sowinski:2015cfa} explored the DCE throughout phase transitions, approaching criticality that resembles the scaling profile stemming from fluid flow turbulence of Kolmogorov-type. It also emulated the important concept of informational turbulence \cite{Sowinski:2016vxz}. Since the DCE has been employed in different areas, in particular bringing new aspects of AdS gravity and instabilities regarding several models, the DCE is here aimed to investigate instabilities that lead a spherically symmetric scalar field to undergo gravitational collapse in asymptotically AdS spacetimes},  also embracing the case where the initial scalar field configuration has either a small or a large amplitude, yielding a black hole \cite{Buchel:2012uh}. When one solves the equations of motion derived from Einstein--Hilbert gravity, with a negative cosmological constant, coupled to a scalar field, the AdS spacetime was shown to be unstable, as a consequence of a resonant wave mode coupling, which yields low-frequency modes to spread out energy towards high-frequency wave modes, exciting them \cite{Bizon:2011gg}.
It thus characterizes the so-called weakly turbulent instability,  also inducing the formation of an apparent horizon for arbitrary collections of initial configurations. Turbulence in this context relies on the causal relationship among points in AdS. Light-like signals can land up at the AdS time-like boundary,  within a finite interval of (proper) time. Therefore any initial configuration recoils recurrently off the boundary of AdS, whereas non-linearity intrinsic to gravitational processes relocates energy out of long-wavelength wave modes into short-wavelength ones, yielding turbulence and subsequently forming an event horizon. 

The critical gravitational collapse stemming from a scalar field profile in AdS has been comprehensively addressed in diverse physical circumstances, utilizing several numerical and theoretical methods \cite{Santos-Olivan:2015yok}. This universal phenomenon appears in a myriad of circumstances, in particular including the gravitational collapse of scalar fields in several types of spacetimes. A case of relevance involves a  scalar field with central symmetry, in the strong gravitational field limit approaching collapse and subsequent black
hole formation.  
AdS${}_{d+1}$ is a homogeneous solution of Einstein’s equations with a negative cosmological constant, having maximal symmetry. 
Although cosmological observations dictate that the current value of the cosmological constant has a non-attractive character, it is still possible to emulate a   cosmological constant with negative value by regularizing gravity in the long-range limit \cite{Kuntz:2019lzq}. AdS${}_{d+1}$ has also a conformal boundary at infinity, which has a peculiar conformal structure of a time-like manifold that is diffeomorphic
to $\mathbb{R}\times S^{d-1}$. Since AdS${}_{d+1}$ is endowed with a natural hyperbolic geometry, asymptotically AdS solutions of Einstein's field equations are suitably addressed as an initial-boundary value problem, with asymptotic boundary conditions. It permits one to discuss the dynamics associated with asymptotically AdS initial data. Imposing reflecting constraints to boundary value problem,  non-linear instabilities of AdS${}_{d+1}$ can be suggested by the nonexistence of asymptotic stability regarding (linear) toy models solutions of the Einstein's field equations. The standard Klein--Gordon equation for a scalar field with conformal coupling can be solved with Dirichlet boundary conditions on the conformal boundary, yielding the energy flux of the scalar field to be constant throughout any foliation at constant time. Therefore it precludes non-trivial scalar field solutions to decay to zero when taking the constant time going to infinity \cite{Holzegel:2015swa}.  
The conformal boundary of asymptotically AdS spacetime differs dramatically from that of asymptotically flat spacetimes, this feature being crucial whenever AdS appears in mathematical-physics.
AdS spacetimes become even more relevant after AdS/CFT has been conjectured,  relating gravity in AdS spacetime to a large-$N$ conformal field theory (CFT) at the boundary of AdS  \cite{Witten:1998qj}.  The dynamics of Einstein's equations, describing weakly coupled gravity in an AdS space, rules the corresponding dynamics of the energy-momentum tensor of strongly-coupled quantum field theories on the boundary of AdS  \cite{Bernardo:2018cow}. 
In the low-energy regime, an unexpected relationship between gravity in AdS and hydrodynamics, governed by the Navier-Stokes equations in the boundary, comprises the fluid/gravity correspondence. 
As dual gravity has been a successful apparatus to study dynamics in quantum field theory, in the near-equilibrium approximation, one can suggest using it also to probe the far-from-equilibrium limit. When strongly-coupled quantum field theories are addressed in the far-from-equilibrium regime, the quark-gluon plasma formation in heavy-ion collisions and quantum quenches in cold atom systems are two among various possibilities of investigation. 
A standing out application tackles open questions involving the quark-gluon plasma as a dual theory to thermal AdS for  $d=4$ \cite{Rougemont:2021qyk,Rougemont:2021qyk1}, where AdS black holes play a prominent role in computing transport coefficients. \cite{Bemfica:2017wps,Bemfica:2019knx}. In the gauge/gravity duality, the gravitational collapse process can be described in the boundary CFT as a thermalization phenomenon \cite{dpr13}, as black hole thermodynamics employs dual 2-point correlators that are ruled by the theorem of fluctuation-dissipation \cite{Caron-Huot:2011vtx}. The case $d=3$ is also particularly important, exhibiting gravitational holographic systems that are dual to 2+1 dimensional strongly-coupled CFT describing condensed matter theories,  encompassing strange metals and graphene among other materials,  and ultracold atomic gases as well. In this respect, gauge/gravity
correspondence supports experimental results about quantum criticality in phase transitions, like the one involving a superconductor-insulator phase. Besides, a sort of open problems regards the linear temperature dependence of resistivity at sufficiently high temperatures in materials such as organic conductors and fullerenes \cite{vandeMarel:2003wn}. 

The AdS instability result asserts that certain types of small perturbations in AdS lead to the formation of black holes \cite{Bizon:2011gg}. Evolving  the vacuum
Einstein equations, with appropriate boundary conditions at the conformal infinity, yield black holes to form, given a long time interval. More precisely, one can generate arbitrarily small perturbations to the initial data of AdS spacetime, evolving according to Einstein's field equations with a reflecting boundary condition on the conformal boundary, driving black hole formation. The property that AdS is non-linearly unstable expresses gravitational turbulence. The emergence of black hole domains implies that non-trivial geometric structures unfold when considering small scales. This phenomenon occurs as a byproduct of the non-linearity underlying initial data evolution \cite{Moschidis:2018ruk}, portraying turbulence amongst strongly interacting wave modes \cite{Bizon:2015pfa,Craps:2014jwa,Evnin:2021buq}.  Since gravitational collapse is the holographic dual to thermalization in the strongly-coupled CFT on the boundary, which attains turbulent regimes, it is not unexpected turbulence to manifest in AdS. Ref. \cite{cw19} showed that from the point of view of the holographic principle, the universal character critical collapse can be mapped onto quantum field theoretical systems on the verge of thermalization, which also present universal characteristics.  Also, the phenomenon of equilibration, consisting of a pure quantum state in a low-temperature phase transiting into a thermalized quantum state at 
higher temperature can be described in the dual gravity arena as an emergent process of gravitational collapse yielding black holes \cite{Bhattacharyya:2009uu}. 

Once the numerical methods are implemented, the gravitational
collapse and the associated critical behavior leading to it may be compared to available observational data regarding coalescing binary systems and black hole mergers, addressing the possibility of bosonic stars \cite{Pretorius:2005gq,LIGOScientific:2017vwq,Brihaye:2013hx}. Several open problems, regarding the stability of the gravitational collapse of a scalar field in AdS, can be investigated in the DCE setup. In addition, the results in this work are endorsed by the most relevant ramifications in the recent literature, besides substantiating new properties of the instability leading to gravitational collapse and weak turbulence. As Ref. \cite{Sowinski:2015cfa} showed that phase transitions in Landau--Ginzburg are governed by the DCE critical behavior, demonstrating that near criticality physical systems are engaging in a turbulent state, one of the aims here 
is to investigate the turbulence and critical behavior associated with the collapse of a scalar field in AdS.

This paper is organized as follows: Section \ref{sec2} is dedicated to presenting solutions of Einstein--Hilbert gravity coupled to a massless scalar in AdS, whose
 full non-linear evolution is implemented and discussed within the numerical approximation, consisting of discretizing
the field equations via spectral methods in Section \ref{sec3}. 
Section \ref{sec4} is devoted to introducing the DCE apparatus as an indicator of AdS instability under small perturbations, which is set off by merging resonant wave modes yielding energy to diffuse from low- to high-frequency wave modes, characterizing turbulence. Section \ref{sec6} addresses  additional analysis, conclusions, and perspectives. 
\section{Field equations}
\label{sec2}

The gravitational collapse in an asymptotically AdS $D$-dimensional spacetime can be settled as follows. Let us consider a massless scalar field dynamics 
$\varphi$ in $D=d+1$ dimensions, minimally coupled to gravitation, with a negative cosmological constant $\Lambda_d$. The associated gravitational action reads
\begin{equation}
S=\int d^{D}x\sqrt{-g}\left [ \frac{1}{16\pi G_d} (R-\Lambda_d)-g_{ab}\nabla^a \varphi \nabla^b \varphi\right ],
\end{equation}
where $G_d$ is the Newton constant, $R$ the scalar curvature, $g$ the metric determinant. We focus in spherically symmetric configurations with metric \cite{Bizon:2011gg,dpr13}
\begin{equation}
\!\!\!\!\!\!\!\!{\scalebox{1.04}{$ds^2\!=\!\sec^2\!\left(\frac{x}{\ell}\right)\!\left[-Ae^{-2\delta}dt^2\!+\!\frac{1}{A}dx^2\!+\!\ell^2\sin^2\!\left(\frac{x}{\ell}\right)d\Omega^2\right]$,}}\!\!
\end{equation}
where the squared AdS radius, which is also its characteristic lengthscale, reads $\ell^2=-d(d-1)/2\Lambda_d$, and $d\Omega^2$ denotes the solid angle element of the $(d-1)$-dimensional unit sphere. The areal radial coordinate $r$ can be related to $x$ by the expression  
\beq
\label{areal}
r = \ell \tan(x/\ell).\eeq The functions 
$A(t,x)$, $\delta(t,x)$ and $\varphi(t,x)$ are defined in the range $x\in(0,\pi/2)$. The standard vacuum AdS corresponds to taking $A=1$, $\varphi=0=\delta=0$. 
In what follows the prime [dot] denotes the derivative of the quantity with respect to $x$ [$t$].
When the ancillary variables \beq
\Phi(t,x)&=&\varphi'(t,x),\\
\Pi(t,x)&=&A^{-1}(t,x)e^{\delta(t,x)}\dot\varphi(t,x),\label{pif}\eeq
are taken into account, the field equations read  \cite{dpr13}
\begin{eqnarray}
A'&=&(1-A)\frac{(d-2\cos^2(x/\ell))}{\cos(x/\ell)\sin (x/\ell)}+\delta'A,\label{ee2}\\
\delta'&=&\frac{4\pi G_d\ell}{1-d} \sin(2x/\ell) (\Phi^2+\Pi^2), \label{ee1}\\
\dot\Pi&=&\cot^{d-1}(x/\ell)\left(\tan^{d-1}(x/\ell) \,e^{-\delta}A\Phi\right)',\\ \label{ee4}
\dot\Phi&=&(e^{-\delta}A\Pi)'. \label{ee3}
\end{eqnarray}
Eq. (\ref{ee4}) is precisely  the Klein--Gordon equation. In what follows $d$-dimensional natural units $8\pi G_d=d-1$ are employed, together with $\ell=1$, for conciseness.

The mass aspect function, encoding the mass-energy content into a radius $x$,  at a given fixed instant $t$,  is given by
\beq
m(t,x)= \frac 1 2 (1-A)\sin^{d-2}(x)\sec^d(x),
\eeq
whereas the Arnowitt--Deser--Misner (ADM) mass is conserved and can be obtained by evaluating the mass aspect function asymptotically, 
\beq
M_{{\scalebox{0.55}{$\textsc{ADM}$}}}=\lim_{x\rightarrow \pi/2} m(t,x).
\label{ADMM}
\eeq 
Eq. (\ref{ee2}) can be equivalently expressed by  \cite{Buchel:2012uh}
\beq
m' = \frac{1}{2} A \tan^{d-1}(x)(\Phi^2+\Pi^2),
\eeq
yielding the ADM mass to be written as
\begin{equation}
M_{{\scalebox{0.55}{$\textsc{ADM}$}}} = \int_0^{\pi/2} \lambda(t,\xi)\, d\xi ,
\label{ADMM_int}
\end{equation}
for 
\beq
\lambda(t,x)=\frac{1}{2}\tan^{d-1}(x) \,A(t,x)\,\left(\Phi^2(t,x)+\Pi^2(t,x)\right),\label{lambda}
\eeq
which can be interpreted as a linear energy density to be used in the DCE calculation,
instead the volumetric energy density $\rho(t,r)$ given in Section \ref{sec4}. Here given the volumetric energy density $\rho(t,r)$ and Eq. (\ref{areal}), the linear energy density 
\be
\lambda(t,x)=4\pi \tan^2(x) \rho(t,x),
\ee
is prescribed as the main ingredient to compute the DCE, $S[{\scalebox{.85}{\textsc{$\lambda$}}}]$, in Section \ref{sec4}.

One can solve  the coupled system of field equations (\ref{ee2}) - (\ref{ee4}) using a code akin to the used in Ref. \cite{dpr13}, utilizing the following
Gaussian-like initial conditions,
\beq
\Phi(t=0,x)&=&0,\\
\Pi(t=0,x) &=& {\varepsilon} \exp\left(-\frac{\tan^2(x)}{\sigma^2}\right).\label{pi0}
\eeq
It means that the scalar field configuration is spatially-localized, obeying the main prerequisite for computing the DCE thereafter. Besides, it propagates along the time coordinate as a tapered wave packet, which fastly collapses for large amplitudes and shapes an apparent horizon at which $A(t, x)$ attains a null value. 
One can push energy into the physical system when the scalar field is perturbed. We want therefore to establish the relationship between the two cases. The first one consists of unstable domains -- and eventual subsequent turbulence  -- arising from perturbation theory,  given small values of $\varepsilon$. The second case is the confirmation of gravitational collapse of the scalar field leading to the formation of a black hole, using numerical methods for finite values of $\varepsilon$. Numerical evidence about configurations inhabiting stability islands, gripped by periodic solutions, show that the energy transfer between low- and high-frequency wave modes is counterbalanced, yielding a static energy spectrum corresponding to stable solutions. 

\section{Numerical solver}
\label{sec3}
{To solve the coupled system of field equations (\ref{ee2})-(\ref{ee4}) we start up with a version of the {\sc Rio} code \cite{bcdr18,aabd21}, adapted to the problem as studied in Ref. \cite{dpr13}. An extra module is added, developed to deal with the DCE calculations as a post-processing treatment of data. Fundamentally the {\sc Rio} code uses the Galerkin-collocation method \cite{boyd}  to discretize the field equations, which therefore are integrated as a dynamical system using the classic Runge--Kutta method. In this particular setting the evolution was implemented as semi-constrained, that is, we solve the non-linear differential equation 
\be
\dot A(t,x) = - \sin(2x)A^2(t,x) \Phi(t,x) \Pi(t,x)e^{-\delta},
\ee
as part of the dynamical system, updating the spectral modes associated to $\delta$ using Eq. (\ref{ee1}), and monitoring the ADM mass calculated by either Eq. (\ref{ADMM}) or (\ref{ADMM_int}), at each time interval.  The spectral basis constructed satisfies the boundary conditions at $x=0$ and $x=\pi/2$,
particularly for $d=3$ \cite{Bizon:2011gg} and all the results are presented for AdS${}_4$. For other spatial dimensions, the boundary conditions change, and the basis for the numerical solver has to be adapted. This is quite relevant for holographic applications where we have to perform specific numerical fits near the conformal infinity. However, in the spectral approximation, the Chebyshev polynomials are treated in the interval $[-1,1]$, independently of the boundary conditions. Thus, the basic building is feasible and implemented with high quality. Nevertheless, the current version of our solver could give a good idea close to the conformal boundary for $d > 3$, if we change the cutoff at the ultraviolet regime using an $\ell\ne 1$. This will be equivalent to drifting the mirror position.  
%For this work, we developed a {\sc Fortran} solver using LAPACK and BLAS, which are open source and optimized libraries. The additional module uses the optimized {\sc Python} libraries for trapezoidal integration and fast Fourier transform to calculate efficiently the DCE. This last script has been calibrated in a previous work \cite{bd22}. For a typical run with $90$ grid points, and a time step of $10^{-4}$, the {\sc Fortran} solver takes $90$ seconds for the massless scalar field traveling to infinite and go back in $\approx\pi$ units of proper time. Each DCE value calculated with the {\sc Python} script takes $0.77$ seconds. All the runs were performed on a $1.8$ GHz Dual-Core Intel Core i5, under OsX Big Sur.
}

\section{DCE, turbulence and black hole formation}
\label{sec4}
The DCE is a logarithmic measure of information entropy. 
The base used in the logarithm is essentially a convention that better suits the physical system under investigation. Choosing a base-2 logarithm indicates bits as the unit information entropy. In the information processing context, it regards the number of bits necessary to encode the event. When employing the base-$e$ logarithm, one measures information entropy in the natural unit of entropy\footnote{Thus 1 nat quantifies the information for an event with an occurrence probability equaling $e^{-1}$. Depending on the logarithm base employed, one can relate different units of information as 1 nat is equal to 
$(\ln 2)^{-1}$ shannon = $(\ln 10)^{-1}$ hartley.} (nat). 
The lowest information entropy corresponds to a random variable consisting of a single event, with a respective probability equal to one, whereas the highest information entropy regards a set of equiprobable events \cite{gs12a}.
The DCE encodes the average number of bits required to represent a spatially-localized physical configuration, displaying the lower bound for the number of bits that are demanded to encrypt the configuration of the physical system.
To compute the DCE, one must first calculate the Fourier transform of the energy density,  
\begin{eqnarray}
{\scalebox{.87}{\textsc{$\rho$}}}({k}) =({2\pi})^{-1/2} \int_0^{+\infty} {\scalebox{.87}{\textsc{$\rho$}}}(\mathsf{r})e^{-i{k}\mathsf{r}}d\mathsf{r},\label{ftrans}
\end{eqnarray} The different waves modes correspond to distinct values of the wavenumber and are weighted and calibrated by the modal fraction, defined as 
\begin{eqnarray}\label{modall}
h_{\scalebox{.8}{\textsc{$\rho$}}}(k) = \frac{|{\scalebox{.87}{\textsc{$\rho$}}}({k})|^2}{\int_{-\infty}^{+\infty}|{\scalebox{.87}{\textsc{$\rho$}}}({\mathtt{k}})|^2d\mathtt{k}}.\label{modalf}
\end{eqnarray} 
Therefore one can quantify the information content and the complexity converted in a coded form into the shape of the energy density, associated with the scalar field in AdS, by computing the DCE
\cite{gs12b}, 
\begin{eqnarray}
S[{\scalebox{.87}{\textsc{$\rho$}}}] = - \int_{-\infty}^{+\infty} \accentset{\star}{h}_{{\scalebox{.8}{\textsc{$\rho$}}}}({{\mathtt{k}}})\log\accentset{\star}{h}_{{\scalebox{.8}{\textsc{$\rho$}}}}({\mathtt{k}})\,d\mathtt{k},\label{ce1}
\end{eqnarray}
where $\accentset{\star}{h}_{\scalebox{.8}{\textsc{$\rho$}}}({k})=h_{\scalebox{.8}{\textsc{$\rho$}}}({k})/h_{\scalebox{.8}{\textsc{$\rho$}}}^{\scalebox{.52}{\textsc{max}}}({k})$. Here  $h_{\scalebox{.8}{\textsc{$\rho$}}}^{\scalebox{.52}{\textsc{max}}}({k})$ stands for the maximal value of the modal fraction, corresponding to a determined value of $k$ whose  power spectrum reaches a global maximum. The DCE estimates  the dynamical order of the solutions of the Einstein--Klein--Gordon system. 
The lower the localization of the scalar field,  the lower the DCE is. Also, the stability of the physical system described by the energy density is usually achieved at the global minima of the DCE. At these critical points, the outcome of the scalar field source is the most certain, as biased toward a weighted domain in wavenumber space. Specifically for the case $d=3$ to be hereon studied, the scalar field configuration in  AdS${}_4$ can be characterized by a collection of normal modes ${\sc E}_i$, with corresponding angular frequencies $\omega_i = 2i + 3$. In the wavelength space, the phenomenon of energy diffusion to
high-frequency wave modes is related to the commensurability of the spectrum of frequencies, yielding the non-linear gravitational collapse to be mediated by resonant interaction.
It implies energy readily to be exchanged among the wave 
modes, hence setting in turbulence.

It is worth mentioning that spatial correlations of the energy density can be still quantified by the 2-point correlator, expressed as
\beq
G(r) = 
\int_0^{+\infty}\rho({r}+\mathsf{r})
\rho(\mathsf{r})\,d\mathsf{r}.
\eeq
Therefore the DCE consists of information entropy that underlies correlations. The associated power spectral density, regarding a mode with wavenumber $k$, reads \cite{Gleiser:2018kbq} 
\beq
P(k) \sim \left\| \int_0^{+\infty} e^{i{k}\mathsf{r}}\rho({\mathsf{r}})\,d{\mathsf{r}}\right\|^2,
\eeq 
and identifies spatial fluctuations of the energy density.

\begin{figure}[H]
\begin{center}
\includegraphics[width=1.6in,height=1.7in]{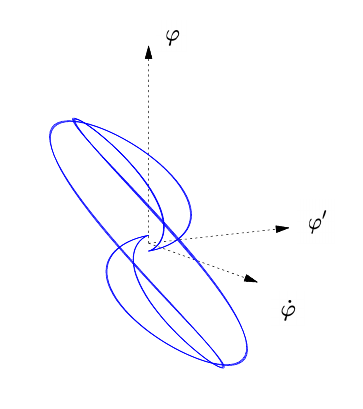}
\includegraphics[width=1.6in,height=1.8in]{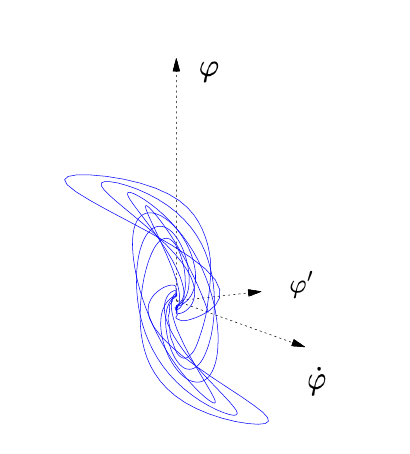}
\includegraphics[width=1.6in,height=1.9in]{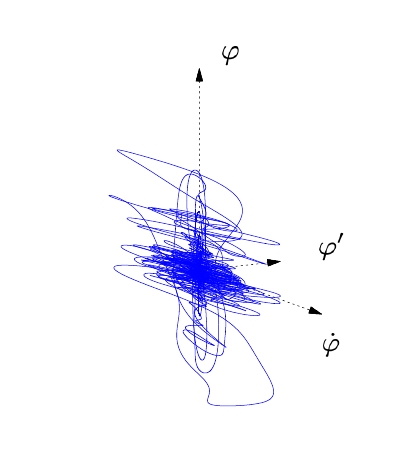}
\caption{{Route to the turbulence for the Einstein--Klein--Gordon system in an asymptotically} AdS$_{d+1}$ {spacetime}, for $d=3$: phase-like diagrams for the scalar field $\varphi$ as a function of  $\dot\varphi$ and $\varphi^\prime$ at $x=0$ up to final time $t_f=20$. The left plot on top   regards $\varepsilon=0.5$, whose dynamics is periodic; the right plot on top corresponds to the choice  $\varepsilon=3.0$ and has a quasi-periodic  dynamics, evincing an incipient unstable and chaotic  phase; the plot on the bottom takes  
$\varepsilon=5.0$, illustrating turbulent dynamics, akin to that by  Ruelle--Takens \cite{nrt78}. These calculations were performed for an expansion in Chebyshev polynomials with 41 terms (collocation points), and replicate the results in Refs. \cite{jrb11,Bizon:2011gg,dpr13,Santos-Olivan:2015yok,ss16b}.} 
%Each temporal series takes approximately  30 minutes of real time calculation  in a processor of 2.4 GHz Intel Core i5.
\label{route}
\end{center}
\end{figure}
\begin{figure}[H]
\begin{center}
\includegraphics[width=2.5in,height=2.2in]{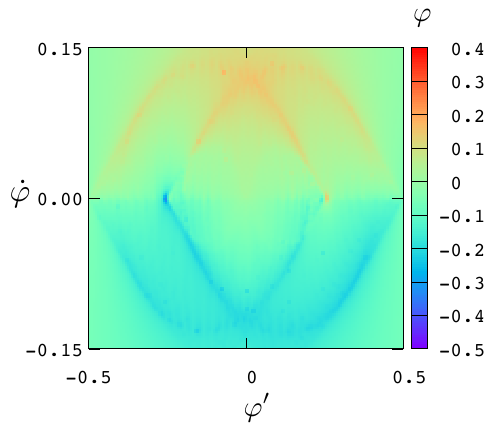}
\includegraphics[width=2.5in,height=2.2in]{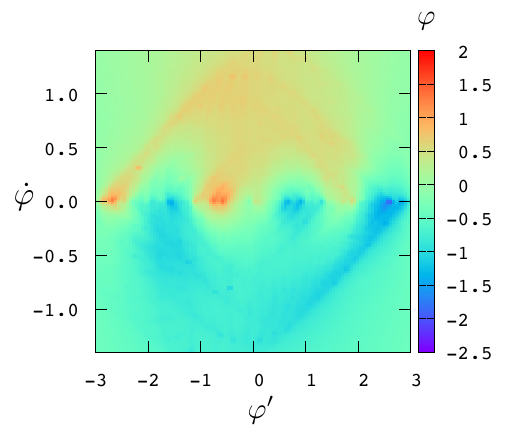}
\includegraphics[width=2.5in,height=2.2in]{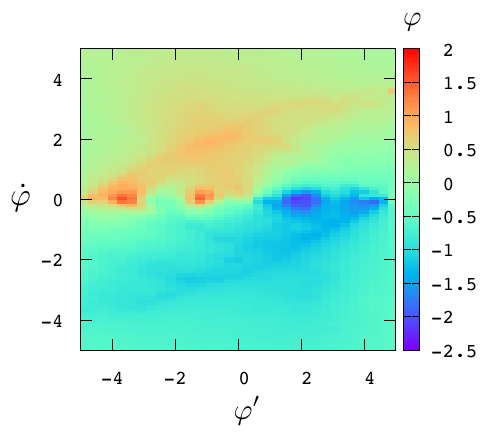}
\caption{Same conditions of Fig. \ref{route} but now $\dot\varphi$,  $\varphi^\prime$  and $\varphi$ (multiplied by $10^2$) are plotted for all $t$ (evolved) and for all $x$ in the domain of $\varphi$. Clearly the periodic structure breaks up at  three crossing times. For $\varepsilon=5.0$, a turbulent regime sets in and a black hole is formed.} 
\label{route_global}
\end{center}
\end{figure}

Fig. \ref{route} makes  explicit the role 
of the perturbation parameter $\varepsilon$ in (\ref{pif}). For an initial small-amplitude $\varepsilon$, the scalar field makes a round trip in $\approx \pi$ units of proper time.
   It came back inverted (downside), that is, when it hits $x=0$ the profile approximates the initial one with amplitude $-\varepsilon$. Then it goes again to infinite and came back upside. The Möbius strip takes $\approx 2\pi$ in proper time. However, this picture vanished with the action of time. There is a critical range for $\varepsilon$  triggering a resonant instability, defining a minimum non-vanishing scalar field amplitude.
We obtain the necessary conditions for such a resonant instability in the gravity-scalar field system, and this analysis can be directly applied to scalar (and other) fields in asymptotically AdS spacetimes.
Complying to formal results in Ref. \cite{Menon:2015oda}, the plot on bottom in Fig. \ref{route}  numerical analysis discloses energy cascading in a turbulent regime, with a scalar field initial amplitude in Eq. (\ref{pif}) fixed by $\varepsilon\sim 5$. Fig. \ref{route_global} generalizes Fig. \ref{route} taking into account the global role of $x$, up to the boundary. Similarly to Fig. \ref{route}, a resonant instability arises, settling in turbulence. 

\begin{figure}[H]
\begin{center}
\includegraphics[width=3.2in,height=1in]{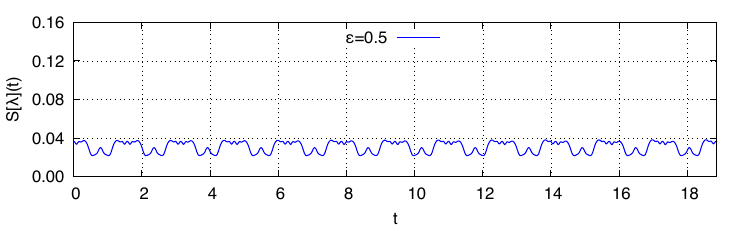}
\includegraphics[width=3.2in,height=1in]{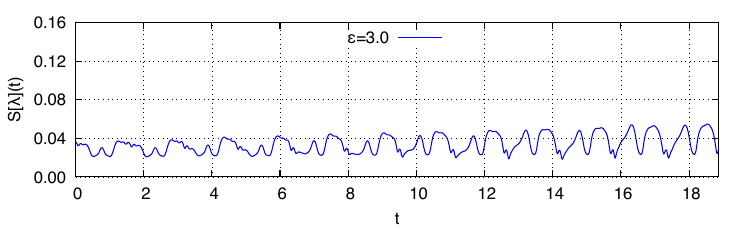}
\includegraphics[width=3.2in,height=1in]{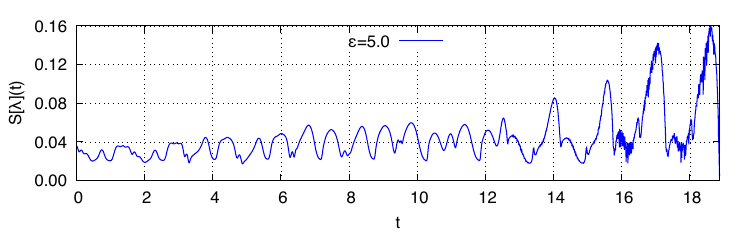}
\caption{{DCE as a function of time. Plot on top for $\varepsilon=0.5$ the dynamics is periodic; in the middle plot for $\varepsilon=3.0$ the dynamics is quasi-periodic, beginning  instability;  plot on bottom for 
$\varepsilon=5.0$ the dynamics is turbulent.  }}
\label{DCE}
\end{center}\vspace{-3.6cm}
\end{figure}

When the DCE is plotted with respect to the proper time, Fig. \ref{DCE} illustrates fundamentally three well-defined regimes of evolution. For the perturbation parameter $\varepsilon=0.5$, there is a periodic dynamics (with a period close to $\pi/2$), whereas  $\varepsilon=3.0$ regards a quasi-periodic, turning into the onset of turbulent after increasing instability, at  
$\varepsilon=5.0$, as a manifestation of the energy cascade into high-frequency modes.  This mechanism can be thought of as being a cascade of quasi-periodic two-fold bifurcations mimicking the torus two-fold bifurcations, in the transition into a weak turbulent state.

\begin{figure}[H]
\begin{center}
\includegraphics[width=3.5in,height=2.5in]{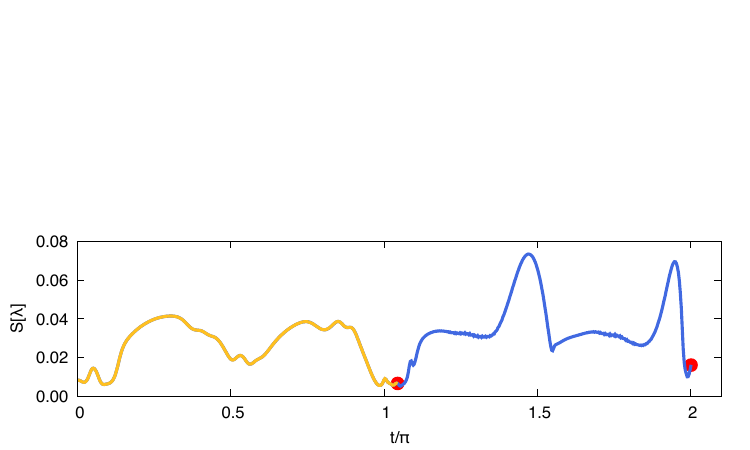}
\caption{{DCE as a function of time, close to one critical amplitude for the initial scalar field. A subcritical evolution with $\varepsilon \approx 7.56$ (golden-blue) a black hole forms near $t\approx 2\pi$ (red point); a supercritical evolution with $\varepsilon \approx 7.57$ (golden) a black hole forms near $t\approx \pi$ (red point).} }
\label{DCE_time}
\end{center}
\end{figure}

The results in Fig. \ref{DCE_time} contemplate the scalar field perturbations in AdS that can lead to a resonant instability associated with a cascade of energy among different frequency wave modes. The scalar field coupled to gravity can be thought of as being a configuration with an infinite number of 
coupled non-linear oscillators. At a finite time interval, 
an appreciable amount of energy is resettled solely to a finite collection of wave modes. For $\varepsilon\approx 7.57$, when the proper time approaches $\pi$, one can expect power-law comportment of the energy-to-mode ratio, near to the gravitational collapse into a black hole. Besides, since the energy of the scalar field is encrypted by the energy density, used to compute the DCE, Fig. \ref{DCE_time} shows that for two precise values of the temporal coordinate, the energy cascades to subsequently smaller
scales, as the gravitational collapse end-stage is a black hole, always in the neighborhood of integer multiples of $\pi$.
 It emulates in the context of the DCE the results in Ref. \cite{Bizon:2011gg}, which showed that when analyzing Einstein's field equations coupled a massless scalar field, the energy transfer to high-frequency modes is not an endless mechanism, as energy clustering onto subsequently smaller
scales unavoidably yields a black hole as a final stage of the gravitational collapse.

\begin{figure}[H]
\begin{center}
\includegraphics[width=3.1in,height=1.1in]{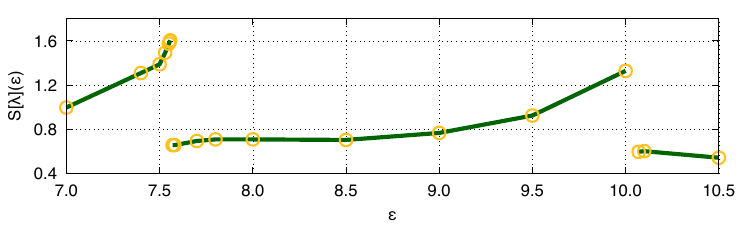}
\includegraphics[width=3.in,height=1.1in]{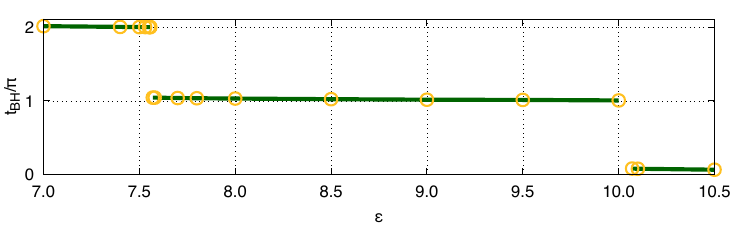}
\caption{{DCE (plot on top, multiplied by $10^2$) and crossing time (plot on the bottom) as a function of initial amplitude.} }
\label{DCE_BH}
\end{center}
\end{figure}
Fig.  \ref{DCE_BH} displays a closer view for the critical gravitational behavior grasped in Fig.  \ref{DCE_time}.
It has been reported multiple critical values for the present context \cite{Santos-Olivan:2015yok}.
We choose one of them, let us say $\varepsilon^{*}$, between $7.56$ and $7.57$. Note that these values may change for each
initial data set, but the behavior seems to be universal. 
For $\varepsilon=7.56$ (subcritical) the scalar field collapses to a black hole in the second trip to infinity for a crossing time of $\approx 2\pi$. For $\varepsilon \lesssim \varepsilon^*$ the DCE grows without saturation, reaching a possible maximum. 
For $\varepsilon=7.57$ (supercritical)  the scalar field collapses to a black hole in the first trip to infinity for a crossing time of $\approx \pi$. It cannot escape for a second round trip. It has been reported with accurate precision the Choptuik’s critical behavior for this situation \cite{Santos-Olivan:2015yok}. On the left branch a type I phase transition occurs (with a mass gap), whereas on the right branch a type II phase transition unveils. The DCE complies with these non-trivial features.  For $\varepsilon \gtrsim \varepsilon^*$ again the DCE grows and reaches a possible maximum near the next critical value.  The DCE falls into the critical transition and looks like a discontinuity. 
The discontinuity of the DCE is portrayed in Fig. \ref{DCE_BH} as a function of the perturbation parameter, in the plot on top. One can realize that both the DCE and the derivative $dS[\lambda](\varepsilon)/d\varepsilon$ are monotonically increasing functions of $\varepsilon$, also being $d^2S[\lambda](\varepsilon)/d\varepsilon^2$ a non-negative monotonic function, for values of the perturbation parameter in the range up to $\varepsilon \approx 7.56$, which matches black hole formation with the DCE equals $\sim$ 0.0161 nat. Then there is a sharp discontinuity of the DCE at $\varepsilon \approx 7.57$, assuming the value $\sim$ 0.0066 nat, reaching $\sim$ 1.3 nat
at $\varepsilon \approx 10.0$, where again the gravitational collapse yields a black hole. There are discontinuities of the DCE corresponding to the very same values of the crossing time, to the perturbation parameter. For all values of the perturbation parameter, a black hole forms, however, only for some critical values of $\varepsilon$ there are discontinuities of the crossing time and the DCE as well.

\section{Conclusions and perspectives}
\label{sec6}

The DCE setup was here investigated in the gravitational collapse of scalar fields in AdS. The DCE was shown to be an indicator of AdS instability under small perturbations. In the wavelength space, the DCE quantifies the process in which resonant wave modes combine and make the energy scatter from low- onto high-frequency modes, representing the onset of turbulence. 
From the DCE point of view, energy and momentum are transported throughout the wavenumber space among different modes. 
From the dual hydrodynamical description, the derived results make it correspond to the formation of fluid vortexes in distinct length scales at turbulent stages. 
\clt{The emergence of strong coherent vortex structures can dominate the fluid flow after some time elapsing. These vortexes comes up as fluctuations that are anomalous at small scales, and can present a lifetime that is much longer than their characteristic swirl turnover time, emulating some of the results in Ref. \cite{Carrasco:2012nf} for the $d=3$ case. With the  dynamical evolutions, two vortexes might combine and merge into a single one, as long as they have angular momentum  with the same orientation, creating  a larger scale vortex. As this processes conserves energy, it plays the role of a mechanism driving inverse energy cascade. 
The advent of vortexes and their  dynamics nevertheless constitute a reasonable number of intricate questions, being the initial data an important feature about the critical behavior. }

The kinetic energy of the fluid flow under turbulence resides in large-scale structures and the energy cascades from them to small-scale ones by inviscid and inertial channels. Still, in the context of the DCE, the dual turbulent fluid flow can be grasped as a superposition of a spectrum of flow fluctuations among the wave modes that constitute the scalar field solutions and turbulent vortexes. 
 The initial condition (\ref{pi0}), reproducing a  Gaussian-like pulse with temporal and spatial evolution by the field equations,  is widened when a range in the variable space $(t,x)$ is considered. There is a critical amplitude of the scalar field velocity (\ref{pif}), regulated by the perturbation parameter $\varepsilon$ of the initial boundary condition (\ref{pi0}), above which gravitational collapse is detected. Below it, a non-linear
field dissipation mechanism engages in the gravitational collapse, ruling it out. In addition, as usually implemented in field theory, one can define the particle number of the scalar field as a function of the positive angular frequency and the scalar field amplitude %, given by $N=\sum_i 4\omega_i\varepsilon^2$ 
\cite{Buchel:2014xwa}. Here the DCE provides a supplementary hint regarding the turbulence increment. 
The conservation of energy and the particle number prevents the whole energy to be transferred to high-frequency wave modes, which present higher energy weighted to each mode. In such a way,  any quota of additional energy implies the wave modes population to lower. Therefore, for the particle number to be conserved while energy is being injected into high-frequency wave modes, the system configuration should be more occupied with low-frequency wave modes. It implements the concept of dual cascading process in AdS${}_4$, and one may speculate the role played by the duality between gravity in AdS${}_4$ and hydrodynamics governed by the Navier--Stokes equations on the AdS${}_4$ boundary. Conformal relativistic hydrodynamics can describe inviscid incompressible fluid flows in (2+1)-dimensional fluid
dynamics, wherein the integral of the vorticity (squared) is a  conserved quantity as well as the energy. This system is dual to asymptotically AdS${}_4$ black holes, with non-vanishing shear viscosity-to-entropy density radio that subleads the 
temperature of the black hole. Therefore, an inviscid approximation holds at finite temperature, implying that perturbations in asymptotically AdS${}_4$ yield turbulence as a consequence of a cascade of energy to large-scale structures, in full compliance to the analysis of Navier--Stokes fluid flows.

The black hole apparent horizon can be approached for the amplitude of the scalar field related to the initial profile. The horizon is formed after a definite number of recoils off the boundary of AdS boundary, also encompassing the straight collapse  \cite{Santos-Olivan:2015yok}. 
%In fluid/gravity correspondence, the gravitational collapse here may be transliterated into fluid dynamics in the turbulent regime, evoking wave turbulence. 
One may argue whether the gravitational collapse of a scalar field in more general conditions can lead to different turbulent profiles, including the possibility of the DCE to describe the holographic account of  phase transitions in 
strongly-coupled quantum systems
\cite{Brihaye:2019dck}. 
The method used here can be forthwith emulated in AdS${}_5$, for the family of generalized black brane solutions. From the general behavior of these solutions, one can expect it to give an account of the dual description of thermalization in models with confinement. Indeed, when switching on a scalar field of amplitude $\varepsilon$, it yields an infalling planar shell toward the interior of AdS, leading to a black brane formation and thermalization in the dual CFT \cite{Craps:2015upq}. Finally, some of the quantum aspects of AdS instabilities were addressed in Ref. \cite{Kuntz:2017pjd} and may be refined in the light shed by the results heretofore. 
\clt{Ref. \cite{Garfinkle:2011tc} points to the  formation of an apparent horizon as a function of the parameters characterizing an
initial profile of a massless scalar field in AdS. It provided additional features about 
thermalization processes when the gravitational dual system is approached. Besides, one can explore scalar field with a potential. Finally, Ref. \cite{Garfinkle:2011hm} suggested the black hole formation as the emergence of an apparent horizon, for sufficiently high values of the scalar field amplitude. Using the time elapsed to the black hole  formation to estimate thermalization in field theory, 
thermalization was shown to set in fastly, near the causal bound.}

\paragraph*{Declaration of competing interest.} The authors declare that they have no known competing financial interests or personal relationships that could have appeared to influence the work reported in this paper.

\paragraph*{Data Availability Statements:}  the datasets generated during and/or analyzed during the current study are available from the corresponding author upon reasonable request.

\subsubsection*{Acknowledgments}
WB thanks Henrique de Oliveira and Leopoldo Pando--Zayas for discussions about some correlated issues studied in this work.  
RdR~is grateful to FAPESP (No. 2021/01089-1 and No. 2022/01734-7) and the National Council for Scientific and Technological Development -- CNPq (Grants No. 303390/2019-0), for partial financial support.

\appendix

\begin{figure}[H]
\begin{center}
\includegraphics[width=2.3in,height=2.5in]{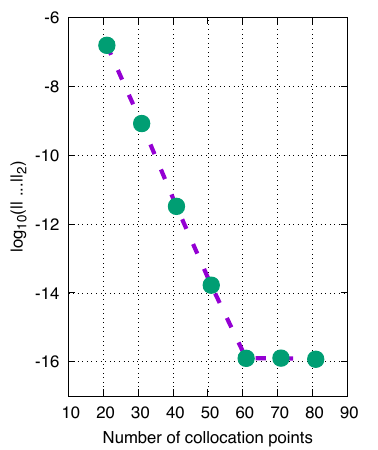}
\caption{\textcolor{black}{Exponential convergence of  $||\Pi_0(r)-\Pi_N(0,r)||_2$ for the initial data given by Eq. (\ref{pi0}), for $\varepsilon=1/2$ and $\sigma=1/4$. This convergence is representative for other parameter choices.}}
\label{initial_data}
\end{center}
\end{figure}

\textcolor{black}{\section{Numerical error analysis}
In order to certificate that our numerical code is robust, first an error measure is defined by means of a root mean square (RMS) norm $||f||_2$,
\begin{equation}
||f||_2=\left\{\frac{1}{2}\int_{-1}^{1}|f|^2 dy\right\}^{1/2}.
\label{eq:L2}
\end{equation}
where $|f|$ is an expected computational zero. We calculate the spectral modes to numerically reconstruct the initial data given by Eq. (\ref{pi0}). Thus $f=\Pi(0,x)-\Pi_N(0,x)$, where the subscript $N$ stands for the numerical reconstruction of the initial data. The numerical integration is performed using a trapezoidal  formula with ten times more points than the correspondent number of collocation points $n$. Figure \ref{initial_data} displays the exponential convergence for the initial data, as expected, up to the working precision. The convergence clearly saturates for $n>60$.}
\begin{figure}[H]
\begin{center}
\includegraphics[width=3.5in,height=1.2in]{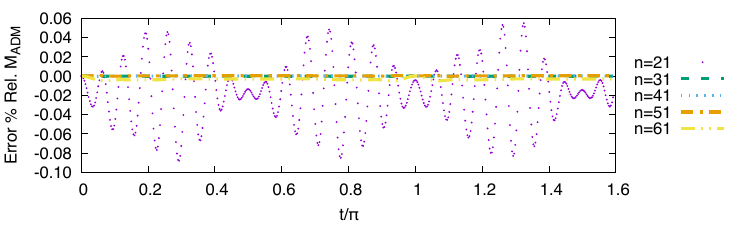}
\caption{\textcolor{black}{Percent relative error of the ADM mass as a function of time. The evolved initial data has the same parameters as used in Fig. \ref{initial_data}. This convergence is representative for other parameter choices.}} 
\label{admass}
\end{center}
\end{figure}
\begin{figure}[H]
\begin{center}
\includegraphics[width=3.5in,height=1.2in]{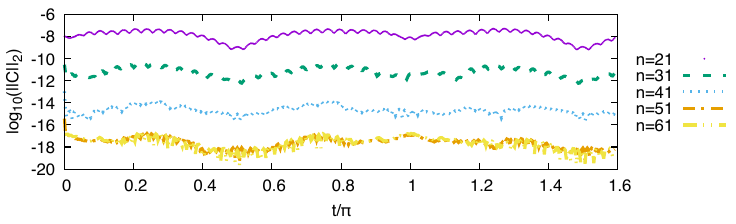}
\caption{Exponential convergence of the constraint error as a function of time. The evolved initial data has the same parameters as used in Fig. \ref{initial_data}. This convergence is representative for other parameter choices.} 
\label{Constraint_t}
\end{center}
\end{figure}
\begin{figure}[H]
\begin{center}
\includegraphics[width=2.3in,height=2.5in]{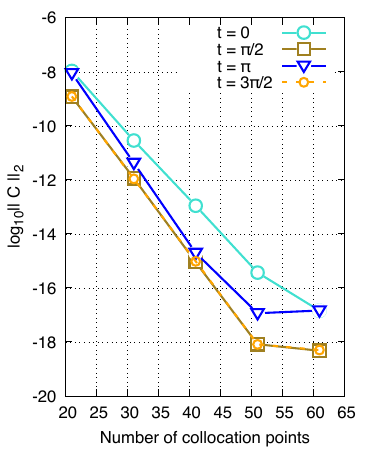}
\caption{\textcolor{black}{Exponential convergence of the constraint error as a function of collocation points number. The evolved initial data has the same parameters as used in Fig. \ref{initial_data}. This convergence is representative for other parameter choices.}} 
\label{Constraint_n}
\end{center}
\end{figure}
\noindent \textcolor{black}{
Also for error analysis, the percent relative error for the ADM mass is defined as
\begin{equation}
\mathcal{E}_{\% r}[M_{{\scalebox{0.55}{$\textsc{ADM}$}}}]=\frac{M_{{\scalebox{0.55}{$\textsc{ADM}$}}}(t)-M_{{\scalebox{0.55}{$\textsc{ADM}$}}}(0)}{M_{{\scalebox{0.55}{$\textsc{ADM}$}}}(0)}\times 100.
\label{eq:relerror}
\end{equation}
Fig. \ref{admass} shows the ADM mass conservation within the numerical error, relative to the initial ADM mass.}

\textcolor{black}{As commented in Section \ref{sec3} the implemented evolution is semi-constrained, thus we have to monitor only the constraint given by Eq. (\ref{ee1}). Defining \beq f=\mathcal{C}=\delta'+2\pi G_d\ell \sin\left(\frac{2x}{\ell}\right) (\Phi^2+\Pi^2),\eeq measuring the difference between the left-hand and the right-hand side of Eq. (\ref{ee1}), we proceed to calculate the constraint error using (\ref{eq:L2}). Fig. \ref{Constraint_t} shows the evolution of the RMS norm for $\mathcal{C}$. The exponential convergence is apparent which is confirmed by means of Fig. \ref{Constraint_n}.}

\textcolor{black}{Finally, for the used numerical spectral method, Galerkin-collocation, by construction the residual equations are satisfied exactly at the collocation points. Thus the issues attended in this appendix let us reassure that our solver is numerically robust.}

\textcolor{black}{\section{Other computational issues}
For this work, we developed a {\sc Fortran} solver using LAPACK and BLAS, open-source and optimized libraries. The additional module uses the optimized {\sc Python} libraries for trapezoidal integration and fast Fourier transform to calculate efficiently the DCE. This last script has been calibrated in a previous work \cite{bd22}. For a typical run with $90$ grid points,  
and a time step of $10^{-4}$, the {\sc Fortran} solver takes $90$ seconds for the massless scalar field traveling to infinite and go back in $\approx\pi$ units of proper time. Each DCE value calculated with the {\sc Python} script takes $0.77$ seconds. All the runs were performed on a $1.8$ GHz Dual-Core Intel Core i5, under OsX Big Sur.}

\end{document}